**Comment on [arXiv:1810.04634](arXiv:1810.04634), "Hole-pocket-driven superconductivity and its universal features in the electron-doped cuprates"**

J.E. Hirsch, Department of Physics, University of California San Diego, La Jolla, CA 92093

### 1. Preface

The findings recently reported in arXiv:1810.04634 by Yangmu Li, W. Tabis, Y. Tang, G. Yu, J. Jaroszynski, N. Barisic and M. Greven, that "clearly point to hole-pocket-driven superconductivity in these nominally electron-doped materials", and point to "a single underlying hole-related mechanism of superconductivity in the cuprates regardless of nominal carrier type", were predicted in the following letter that was submitted to Nature [Scientific Correspondence Section] on March 2, 1989 [1], shortly after the experimental discovery of electron-doped cuprates announced in Nature on January 26, 1989. The journal declined to publish it. Many more papers on this topic were published by F. Marsiglio and the author in the ensuing 30 years from 1989 to the present [2-13]. In omitting to refer to any of these theoretical works, arXiv:1810.04634 presents a misleading picture.

### 2. Letter submitted to Nature (March 2, 1989, revised May 11, 1989)

**Electron-Superconductivity in Oxides?**

Sir - the recent articles on electron-doped oxide superconductors[1-4] all suggest that experiments have already established that conduction in these materials is fundamentally different than in the hole-doped oxides. In particular, Emery[3] claims that in these materials "the supercurrent is carried by electrons rather than holes," and Rice[4] suggests that these materials provide decisive evidence in favor of the t - J model that is particle-hole symmetric. I would like to disagree with these points of view, and predict that oxygen hole carriers will be found in these materials.

We have recently proposed a theory of superconductivity based on the fundamental asymmetry between electron and holes, within which the existence of oxygen hole carriers is a necessary condition for high temperature superconductivity in oxides to occur.[5,6] If it is established that no oxygen hole carriers exist in the electron-doped oxides our theory will be proven wrong. By the same token, if O hole carriers are found it will lend strong support to our theory as it will be an unexpected finding predicted by it.

Emery himself has mentioned that electrons added on $Cu^{++}$ could induce holes on neighboring oxygens. I would like to point out that the very same fact that allows for electron doping of $Cu^{++}$ in these structures, absence of apical oxygen,[1] makes it favorable for O holes to be created as electrons are added. The electron added on $Cu^{++}$ will repel the electrons on neighboring $O^=$ towards other neighboring $Cu^{++}$, and the absence of $O^=$ on top and bottom of the $Cu^{++}$ ions makes it energetically favorable for these $Cu^{++}$ ions to accept the $O^=$ electrons. The net result of adding one electron is thus likely to be several

$Cu^+$ and several O holes. The large direct hopping integrals between oxygens[7] will cause these induced holes to delocalize rather than remain bound to $Cu^+$, and similarly the electrons added to $Cu^{++}$ will delocalize through hopping between Cu and O.

To have both electron and hole carriers at the Fermi surface is the rule rather than the exception in solids. This fact (even without invoking the anisotropy of these materials) can easily explain the negative Hall coefficient observed.[1] For a two-band model with electron and hole carriers the Hall coefficient is given by

$$R = -\frac{1}{n_e e c} \frac{1 - \frac{n_h}{n_e}(\frac{m_h}{m_e})^2}{(1 + \frac{n_h}{n_e}\frac{m_h}{m_e})^2} \quad (1)$$

with $n_e$ ($n_h$) the number of carriers and $\mu_e$ ($\mu_h$) the mobility of electrons (holes). Now we expect the mobility of the hole carriers to be substantially lower than that of the electron carriers. In particular, assuming equal relaxation times for both types of carriers $\mu_e/\mu_h = m^*_e/m^*_h$, with $m^*_e$ ($m^*_h$) the effective masses; both the sizes of the different hopping rnatrix elements involved[7] and many-body effects[5] contribute to yield $m^*_h \gg m^*_e$. Equation (1) then yields a negative Hall coefficient even if substantially more hole than electron carriers exist. We also note that the published Hall coefficient data of Takagi et al.[8] show a sharp drop in the magnitude of the (negative) Hall coefficient as function of doping right before the onset of superconductivity, which suggests from Eq. (1) a sudden increase in the number of hole carriers through the induction process described above. Finally, we recall that even in the "old" oxide superconductor $BaPb_{1-x}Bi_xO_3$ with negative Hall coefficient evidence for heavy hole carriers was found from careful analysis of thermopower data.[9]

There are several spectroscopic experiments that can probe directly the existence of O hole states at the Fermi surface that have been performed for the "hole" oxide superconductors. Hopefully they are being done or will be done soon. We would like to emphasize their importance, since, as argued here, the view[1-4] that the electron-doped oxide materials are electron-carrier oxide superconductors is not a foregone conclusion.

The finding of oxygen hole carriers in the electron-doped oxides will not support all theories where these carriers pair to yield superconductivity. In almost all such theories proposed for the hole-doped oxides the "glue" that pairs the oxygen holes are charge or spin excitations of the $Cu^{++}$ background. In the electron-doped oxides this background is rapidly being destroyed by the added plus induced electrons that turn $Cu^{++}$ into $Cu^+$ a closed-shell ion. Our theory,[5,6] which describes "metallic oxygen" with everything else playing a secondary role, will survive.

J. E. Hirsch
Department of Physics
University of California, San Diego

La Jolla, CA 92093